\documentclass[12pt]{iopart}

\usepackage[T1]{fontenc}
\usepackage{lmodern}
\usepackage{iftex}
\ifPDFTeX
  \usepackage[protrusion=true,expansion=true]{microtype}
\else
  \usepackage[protrusion=true,expansion=false]{microtype}
\fi

\usepackage{graphicx}
\usepackage{here}
\usepackage{color}
\usepackage{wrapfig}
\usepackage{booktabs} 
\usepackage{threeparttable} 
\usepackage{microtype}
\usepackage{sidecap}
\sidecaptionvpos{figure}{c}
\usepackage{hyperref}
\usepackage{siunitx}
\usepackage{bm}
\hypersetup{
    bookmarksnumbered=true,
    bookmarksopen=true,
    hidelinks,          
}

\usepackage{xurl}

\usepackage[
style=phys,
articletitle=false,biblabel=brackets,
chaptertitle=false,pageranges=false,
maxbibnames=3,
minbibnames=1,
url=true
]{biblatex}
\DeclareSourcemap{
  \maps[datatype=bibtex]{
    \map{
      \pertype{article}
      \step[fieldset=url, null]
      \step[fieldset=urldate, null]
    }
  }
}

\AtEveryBibitem{%
  \clearlist{language}
  \clearfield{note}

}

\DefineBibliographyStrings{english}{
    andothers = {\textit{et al\adddot}}
}

\begin{document}

\title["Magneto-optical Kerr effect measurements under bipolar pulsed magnetic fields"]
{Magneto-optical Kerr effect measurements under bipolar pulsed magnetic fields} %Title of the paper

\author{Soichiro Yamane$^{1*}$, Sota Nakamura$^1$, Atsutoshi Ikeda$^1$, \\Kosuke Noda$^2$, Akihiko Ikeda$^2$, and Shingo Yonezawa$^1$ }

\address{$^1$Department of Electronic Science and Engineering, Kyoto University, Kyoto 615-8510, Japan\\
	$^2$Department of Engineering Science, University of Electro-Communications, Tokyo 182-8585, Japan.
}

\vspace{10pt}
\begin{indented}
	\item[]$^*$ E-mail : yamane.soichiro.73f@st.kyoto-u.ac.jp
\end{indented}

\begin{abstract}
	The magneto-optical Kerr effect (MOKE) is a powerful probe of magnetism.
	Its contact-free optical nature makes it potentially well suitable for measurements under pulsed magnetic fields if various difficulties are overcome.
	In this paper, we report the establishment of MOKE measurements under bipolar pulsed magnetic fields up to 13.1~\si{\tesla}.
	% at room temperature using a zero-area-loop Sagnac interferometer.
	The accuracy of the setup was demonstrated by the excellent agreement with static-field results on the $(001)$ surface of a Fe$_3$O$_4$ single crystal.
	Furthermore, clear hysteresis loops of various commercial permanent magnets were successfully observed.
	The capability for rapid characterization of hysteretic properties highlights the versatility of our pulsed-field MOKE setup for both fundamental materials science and engineering applications.
\end{abstract}

%\maketitle

\section{Introduction}

	The magneto-optical Kerr effect (MOKE) is an optical probe of magnetism.
	In most materials, the Kerr angle ($\theta_{\mathrm{K}}$), rotation of the light's polarization, is proportional to magnetization.
	Consequently, ferromagnets such as iron exhibit relatively large $\theta_{\mathrm{K}}$ of the order of milliradians.
	Moreover, MOKE has enabled various discoveries in various materials beyond ferromagnets \cite{Xia2006-rk, Shimano2013-lc, Huang2017-ns}.
	MOKE is therefore a key non-contact probe for exploring frontiers of materials science.

	From an experimental viewpoint, because MOKE measurements requires no electrical contacts, it potentially well matches with pulsed magnetic fields.
	However, only a few MOKE studies under pulsed fields have been reported \cite{Weisheit2006-gt, Chen2013-ll, Lin2017-ro}, owing to limited sample spaces in a pulse-field coil and small $\theta_{\mathrm{K}}$ in most materials.
	Despite these challenges, we recently succeeded in measuring $\theta_{\mathrm{K}}$ of various ferromagnetic or ferrimagnetic samples in pulsed fields exceeding 40~\si{\tesla} and down to 77~\si{\kelvin} \cite{Ikeda2025-en} using a portable pulse-field generator \cite{Ikeda2024-ja} and a zero-area-loop (loop-less) Sagnac interferometer \cite{Xia2006-sg}.
	This accomplishment is driven by our new compact sample fixture matching with a small bore of a coil and a phase-resolved numerical lock-in analysis for MOKE signals \cite{Ikeda2025-en}.
	As a next step, MOKE measurements in bipolar magnetic fields enable the observation of full magnetic hysteresis loops, which is indispensable for characterizing the coercivity and remnant magnetization of magnetic materials.
	These advances further establish pulsed-field MOKE as a promising approach in both fundamental and applied magnetism.

	In this paper, we report successful MOKE measurements under bipolar pulsed magnetic fields up to 13.1~\si{\tesla}.
	The validity and utility of our setup are demonstrated by results from Fe$_3$O$_4$ and various permanent magnets.

\section{Experimental methods}

	For MOKE measurements under bipolar pulsed magnetic fields,
	we used a method essentially the same as our previous work \cite{Ikeda2025-en} for unipolar pulsed fields.
	From the interference signal detected by a photodetector and an oscilloscope, $\theta_{\mathrm{K}}$ is extracted via phase-resolved numerical lock-in analysis \cite{Ikeda2025-en}.
	In this work, we newly developed a flexible command-line interface (CLI) tool for pulsed MOKE measurements \cite{YamaneUnknown-xs}.
	This CLI, mainly built by the Rust programming language, provides efficient and user-friendly tools for analyzing large datasets obtained in pulsed-field MOKE measurements.
	For instance, the core numerical lock-in processing of $10$ million data points is computed in less than $200$ \si{\milli \second} on a standard laptop.
	We used an integration time of $\tau = 1.72$~\si{\micro \second} in our numerical lock-in analysis, and further smoothed the data using a moving average with a window size of $20$ points (corresponding to $10.0$~\si{\micro \second}) to improve the \mbox{signal-to-noise ratio.}

	Bipolar pulsed magnetic fields up to $13.1$~\si{\tesla} were generated using a portable pulse-field generator \cite{Ikeda2024-ja}, which was modified by adding a diode antiparallel to the thyristor to allow a bipolar current oscillation of a single period.
	The magnetic field was measured by monitoring the current flowing through the main coil using a Rogowski sensor.
	The optical configuration of our polar MOKE setup was described in Ref.~\cite{Ikeda2025-en}; the incident and reflected lights are collinear to the field, and the positive direction of magnetic fields is defined as antiparallel to the incident light (Faraday geometry); $\bm{H} \parallel -\bm{k}$, where $\bm{H}$ is the magnetic field and $\bm{k}$ is the wave vector of the incident light.
	We also performed static-field MOKE measurements using a commercial superconducting magnet (Quantum Design, PPMS), employing the same optical geometry as the pulsed-field setup.

	Demonstration of polar MOKE measurements in bipolar pulsed fields was conducted on the $(001)$ surface of a commercial Fe$_3$O$_4$ single crystal.
	Commercial permanent magnets made of various ferromagnetic materials were used to investigate the hysteretic magnetic properties.

	% \section{Results and discussion}
\section{Demonstration of bipolar pulsed-field MOKE measurements using \texorpdfstring{Fe$_3$O$_4$}{Fe3O4}}

	\begin{figure}[ht]
		\centering
		\includegraphics[clip,width=440pt, bb=0.000000 0.000000 439.683371 157.590007]{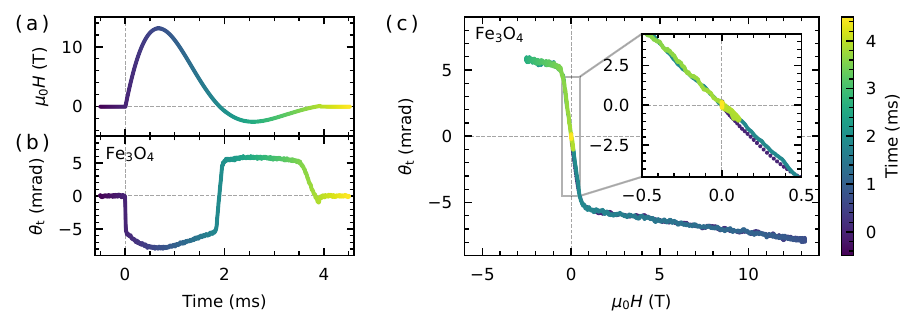}

		\caption{\label{F1}
			Results of MOKE measurements on the $(001)$ surface of a Fe$_3$O$_4$ single crystal under a bipolar pulsed magnetic field up to 13.1~\si{\tesla} at room temperature.
			Time dependence of (a) the magnetic field and (b) the total magneto-optical (MO) angle. (c) $\theta_{\mathrm{t}}$ as a function of magnetic field.
			The inset shows a magnified view around zero magnetic field.
			For all panels, the color of data points indicate the time evolution from the beginning of the pulse, as shown in the colorbar on the right.
		}
	\end{figure}

	Figure~\ref{F1} shows results of a MOKE measurement on a $(001)$ surface of a Fe$_3$O$_4$ single crystal under a bipolar pulsed magnetic field at room temperature.
	Figure~\ref{F1}(a) shows the time dependence of the magnetic field.
	After the field pulse is triggered, the magnetic field reaches $13.1$~\si{\tesla} at $0.7$~\si{\milli \second}, then decreases to $-1.9$~\si{\tesla} at $2.6$~\si{\milli \second}, and returns to $0$~\si{\tesla} at $4.0$~\si{\milli \second}.
	Figure~\ref{F1}(b) exhibits the time dependence of the total magneto-optical (MO) angle ($\theta_{\mathrm{t}} = \theta_{\mathrm{K}} + \theta_{\mathrm{bg}}$), where $\theta_{\mathrm{bg}}$ is a background mainly originating from the Faraday effect of optical components.
	Upon applying the bipolar pulsed magnetic field, $\theta_{\mathrm{t}}$ clearly responds to the time profile of the magnetic field.
	The steep changes between $\theta_{\mathrm{t}} = \pm 5$~\si{\milli \radian} are due to the alignment of ferrimagnetic domains in Fe$_3$O$_4$,
	whereas the gradual changes beyond $\pm 5$~\si{\milli \radian} are attributed to the background Faraday effect as discussed below.
	The magnetic-field dependence of $\theta_{\mathrm{t}}$ is plotted in Fig.~\ref{F1}(c).
	The inset magnifies the region around zero magnetic field.
	Within our experimental resolution, negligible hysteresis loop was observed, consistent with the known soft ferrimagnetic nature of Fe$_3$O$_4$ at room temperature \cite{Heider1987-zq}.

\begin{SCfigure}[\sidecaptionrelwidth][ht]
		\centering
		\includegraphics[clip,width=220pt, bb=0.000000 0.000000 215.835499 204.148004]{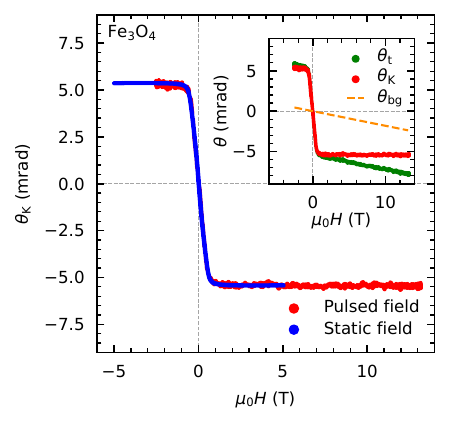}
		\caption{\label{F2}
			$\theta_{\mathrm{K}} (H)$ of a $(001)$ surface of a Fe$_3$O$_4$ single crystal at room temperature.
			Red points present results obtained in the bipolar pulsed magnetic field up to 13.1~\si{\tesla}, while blue points show results measured in a static magnetic field.
			The inset describes the separation of $\theta_{\mathrm{K}}$ (red points) and $\theta_{\mathrm{bg}}$ (orange dashed line) from $\theta_{\mathrm{t}}$ (green points).
		}
	\end{SCfigure}

	To extract $\theta_{\mathrm{K}}$ from the total angle $\theta_{\mathrm{t}}$, the background Faraday rotation $\theta_{\mathrm{bg}} = \alpha H$ is subtracted as illustrated in the inset of Fig.~\ref{F2}.
	The proportionality constant $\alpha$ is determined by fitting the data above 5 \si{\tesla} to the function $\theta_{\mathrm{t}} (H) = \theta_{\mathrm{K}}^{\mathrm{sat}} + \alpha H$, where $\theta_{\mathrm{K}}^{\mathrm{sat}}$ denotes the saturated Kerr angle.
	Figure~\ref{F2} shows the obtained $\theta_{\mathrm{K}} (H)$.
	For comparison, the $\theta_{\mathrm{K}}$ measured in a static magnetic field using a commercial cryostat is also plotted.
	The two datasets show excellent agreement, validating the accuracy of our bipolar pulsed-field MOKE measurements, even under the challenging conditions of rapidly changing magnetic fields with only a duration of a few milliseconds.
	$\theta_{\mathrm{K}}^{\mathrm{sat}}$ is evaluated to be $-5.39 \pm 0.01$~\si{\milli \radian}, which is close to the literature values at room temperature \cite{Simsa1980-us, Fontijn1997-hr}, as summarized in Table \ref{tab:kerr_comparison}.
	The small variations in $\theta_{\mathrm{K}}^{\mathrm{sat}}$ can be attributed to differences in surface orientation and treatment, as well as the use of a broadband incoherent light source in our experimental setup.
	Overall, our MOKE measurements successfully capture the magnetization behavior of Fe$_3$O$_4$, demonstrating the effectiveness of our approach for studying magnetic materials under bipolar pulsed fields.
	\begin{table}[ht]
		\centering
		\begin{threeparttable}
			\caption{Comparison of $\theta_{\mathrm{K}}^{\mathrm{sat}}$ of Fe$_3$O$_4$ at room temperature. All literature values were obtained at a photon energy of $0.8$~\si{\electronvolt} ($1550$-nm wavelength).}
			\label{tab:kerr_comparison}
			\small
			\begin{tabular}{llcl}
				\toprule
				$\theta_{\mathrm{K}}^{\mathrm{sat}}$ (mrad)  & Surface Treatment             & Field                     & Ref.                   \\
				\midrule
				$\mathbf{-5.39 \pm 0.01}$                    & \textbf{As-polished $(001)$}  & $\bm{5-13.1}$\textbf{ T}  & \textbf{This work}     \\
				$-5.14$                                      & Heat-treated $(110)$          & $1$ T                     & \cite{Simsa1980-us}    \\
				$-4.91$                                      & As-polished $(110)$           & $1$ T                     & \cite{Simsa1980-us}    \\
				% $-4.54$ (est.)                               & Annealed                      & Calibrated\tnote{a}  & \cite{Zhang1981-zd}    \\
				% $-4.30$ (est.)                               & As-grown                      & Calibrated\tnote{a}  & \cite{Zhang1981-zd}    \\
				$-4.36$                                      & Annealed $(110)$              & $1.76$ T                  & \cite{Fontijn1997-hr}  \\
				\bottomrule
			\end{tabular}
			% \begin{tablenotes}
			% 	\small
			% 	\item[a] Values estimated by calibrating the applied field ($\sim 0.5$ T) against saturation magnetization data ($\sim 0.6$ T).
			% \end{tablenotes}
		\end{threeparttable}
	\end{table}

\section{Bipolar pulsed-field MOKE measurements on various permanent magnets}

	\begin{figure}[t]
		\centering
		\includegraphics[clip,width=440pt, bb=0.000000 0.000000 445.964018 341.749065]{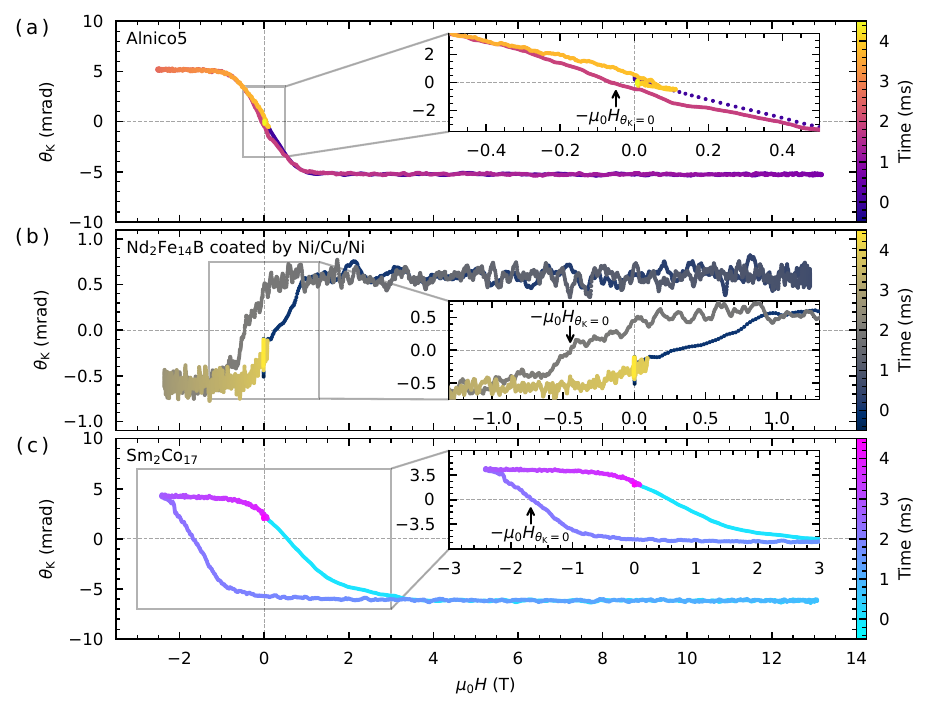}

		\caption{\label{F3}
			Results of MOKE measurements on various permanent magnets under bipolar pulsed magnetic fields at room temperature.
			Hysteresis loops of (a) alnico5, (b) Nd$_2$Fe$_{14}$B with a Ni/Cu/Ni protective coating, and (c) Sm$_2$Co$_{17}$.
			Because details of the starting trajectories vary depending on the magnetic history, we here plot representative data exhibiting largest hysteresises. We note that the overall hysteresis loop shapes themselves represent the intrinsic and reproducible properties of each magnet.
		}

	\end{figure}

	Figure~\ref{F3} displays results of MOKE measurements on (a) an alnico magnet (alnico5, material grade SAN52 \cite{NeoMagUnknown-ms}), (b) a neodymium magnet (Nd$_2$Fe$_{14}$B, material grade N52 \cite{NeoMagUnknown-ms}) with a Ni/Cu/Ni protective coating, and (c) a samarium-cobalt magnet (Sm$_2$Co$_{17}$, material grade SS28 \cite{NeoMagUnknown-ms}) under bipolar pulsed magnetic fields at room temperature.
	Clear hysteresis loops were observed in all samples, highlighting the capability of our bipolar pulsed-field MOKE setup to capture magnetic hysteresis and domain properties.
	We note that magnetic properties of the neodymium magnet were measured through its Ni/Cu/Ni coating. Thus the values of $\theta_{\mathrm{K}}$ originate from the coating layer, whereas the wide hysteresis captures the magnetic properties of the magnet.
	In engineering viewpoints, it is important that hysteretic properties be obtainable from as-bought samples without removing the coating.
	The Ni/Cu/Ni-coated Nd$_2$Fe$_{14}$B magnet exhibits a positive $\theta_{\mathrm{K}}$ at a positive field, in contrast to the other two magnets.
	This positive rotation is consistent with the known positive $\theta_{\mathrm{K}}$ of Ni at $0.8$~\si{\electronvolt} \mbox{under a positive magnetic field \cite{Buschow1983-fs, Delin1999-ov}.}

	To quantify properties as permanent magnets, characteristic fields ($\mu_0 H_{\theta_{\mathrm{K}}=0}$), where $\theta_{\mathrm{K}}$ becomes zero, were determined from the negative-field side of the hysteresis loops.
	This approach ensures reliable values, free from transient effects arising from the initial magnetization process near zero field and the rapid initial field-sweep rate.
	The obtained $\mu_0 H_{\theta_{\mathrm{K}}=0}$ values are approximately (a) 0.05~\si{\tesla} for the alnico5 magnet, (b) 0.45~\si{\tesla} for the Nd$_2$Fe$_{14}$B magnet, and (c) 1.68~\si{\tesla} for the Sm$_2$Co$_{17}$ magnet.
	According to the supplier's datasheets \cite{NeoMagUnknown-ms}, the standard intrinsic coercive fields ($\mu_0 H_{\mathrm{c}j}$) are (a) 0.07~\si{\tesla}, (b) $\geq 1.1$~\si{\tesla}, and (c) $\geq 1.8$~\si{\tesla}, respectively.
	Thus, the measured $\mu_0 H_{\theta_{\mathrm{K}}=0}$ values are lower than the datasheet specifications for all samples.
	This fact can be attributed to the local probing nature of polar MOKE, which is primarily sensitive to the out-of-plane component of the local magnetization ($\bm{M} \parallel \bm{k}$), and is insensitive to in-plane local magnetization ($\bm{M} \perp \bm{k}$).
	When the magnetic field is close to $\mu_0 H_{\mathrm{c}j}$, the magnetization reversal process involves the formation of magnetic domains with various orientations.
	In particular, near surfaces perpendicular to the external magnetic field, domains with in-plane magnetization should tend to be formed so that the stray field outside the sample is minimized.
	Consequently, the out-of-plane component of the magnetization within the light spot for MOKE measurements should decrease more rapidly than the total magnetization measured by bulk methods,
	resulting in $H_{\theta_{\mathrm{K}}=0}$ lower than $H_{\mathrm{c}j}$.

\section{Conclusion}
	We have successfully demonstrated MOKE measurements under bipolar pulsed magnetic fields up to 13.1~\si{\tesla} at room temperature using a zero-area-loop Sagnac interferometer.
	Excellent agreement between pulsed-field and static-field MOKE measurements was achieved on the $(001)$ surface of Fe$_3$O$_4$, validating the performance of our approach.
	The saturated $\theta_{\mathrm{K}}$ obtained from the pulsed experiment was $-5.39 \pm 0.01$~\si{\milli \radian}, close to literature values.
	Moreover, clear hysteresis loops were observed in various permanent magnets, alnico5, Ni/Cu/Ni-coated Nd$_2$Fe$_{14}$B, and Sm$_2$Co$_{17}$.
	The successful observation of distinct hysteresis loops establishes our bipolar pulsed-field MOKE setup as a versatile tool for both fundamental research and rapid characterization in applied magnetism, paving the way for future exploration of novel magnetic materials.

\section*{Acknowledgments}
	% \footnotesize
	This work was supported
	by Grant-in-Aids for Academic Transformation Area Research (A) ``1000 Tesla Science'' (KAKENHI Grant Nos.~JP23H04859 and JP23H04861) from the Japan Society for the Promotion of Science (JSPS),
	by Grant-in-Aids for Scientific Research (KAKENHI Grant Nos.~JP24H00194, JP23K17670, JP22H01168) from JSPS,
	by Japan Science and Technology Agency (JST) FOREST program (Grant No.~JPMJFR222W),
	% by ISHIZUE 2023 of Kyoto University Research Development Program,
	by Iketani Science and Technology Foundation (Grant No.~0361078-A),
	and by The Mitsubishi Foundation (Grant No.~202410051).

\section*{References}
	\printbibliography[heading=none]

@ARTICLE{Weisheit2006-gt,
  title        = {Magnetization reversal of highly coercive {FePt} examined with
                  pulsed microcoils},
  author       = {Weisheit, M and Bonfim, M and Grechishkin, R and Barthem, V
                  and Fahler, S and Givord, D},
  journaltitle = {IEEE Trans. Magn.},
  publisher    = {Institute of Electrical and Electronics Engineers (IEEE)},
  volume       = {42},
  issue        = {10},
  pages        = {3072--3074},
  date         = {2006-10},
  doi          = {10.1109/tmag.2006.880146},
  issn         = {0018-9464,1941-0069},
  url          = {http://dx.doi.org/10.1109/TMAG.2006.880146},
  urldate      = {2025-12-01},
  language     = {en}
}

@ARTICLE{Chen2013-ll,
  title        = {The measurement of magneto-optical Kerr effect of ultrathin
                  films in a pulsed magnetic field},
  author       = {Chen, Xi and Qian, Xuan and Meng, Kangkang and Zhao, Jianhua
                  and Ji, Yang},
  journaltitle = {Measurement},
  publisher    = {Elsevier BV},
  volume       = {46},
  issue        = {1},
  pages        = {52--56},
  date         = {2013-01},
  doi          = {10.1016/j.measurement.2012.05.014},
  issn         = {0263-2241,1873-412X},
  url          = {https://www.sciencedirect.com/science/article/pii/S0263224112002230},
  language     = {en}
}

@ARTICLE{Lin2017-ro,
  title        = {Magneto optical Kerr effect measurement with pulsed magnetic
                  field},
  author       = {Lin, Sih-Ying and Liao, Yi-Kai and Wu, Chii-Bin},
  journaltitle = {Chin. J. Phys.},
  publisher    = {Elsevier BV},
  volume       = {55},
  issue        = {3},
  pages        = {698--702},
  date         = {2017-06},
  doi          = {10.1016/j.cjph.2017.04.010},
  issn         = {0577-9073,2309-9097},
  url          = {http://dx.doi.org/10.1016/j.cjph.2017.04.010},
  language     = {en}
}

@ONLINE{NeoMagUnknown-ms,
  author    = {{NeoMag}},
  booktitle = {NeoMag},
  url       = {https://www.neomag.jp/products_navi/before_catalog.html},
  urldate   = {2025-12-01},
  language  = {ja}
}

@ARTICLE{Delin1999-ov,
  title        = {Calculated magneto-optical properties of cubic and tetragonal
                  Fe, Co, and Ni},
  author       = {Delin, Anna and Eriksson, Olle and Johansson, Börje and
                  Auluck, Sushil and Wills, J M},
  journaltitle = {Phys. Rev. B},
  publisher    = {American Physical Society},
  volume       = {60},
  issue        = {20},
  pages        = {14105},
  date         = {1999-11-15},
  doi          = {10.1103/PhysRevB.60.14105},
  url          = {http://dx.doi.org/10.1103/PhysRevB.60.14105},
  urldate      = {2025-12-01},
  language     = {en}
}

@ARTICLE{Buschow1983-fs,
  title        = {Magneto-optical properties of metallic ferromagnetic materials},
  author       = {Buschow, K H J and van Engen, P G and Jongebreur, R},
  journaltitle = {J. Magn. Magn. Mater.},
  publisher    = {Elsevier BV},
  volume       = {38},
  issue        = {1},
  pages        = {1--22},
  date         = {1983-08-01},
  doi          = {10.1016/0304-8853(83)90097-5},
  issn         = {0304-8853,1873-4766},
  url          = {http://dx.doi.org/10.1016/0304-8853(83)90097-5},
  urldate      = {2025-12-01},
  language     = {en}
}

@ARTICLE{Heider1987-zq,
  title        = {Magnetic properties of hydrothermally recrystallized magnetite
                  crystals},
  author       = {Heider, F and Dunlop, D J and Sugiura, N},
  journaltitle = {Science},
  publisher    = {American Association for the Advancement of Science (AAAS)},
  volume       = {236},
  issue        = {4806},
  pages        = {1287--1290},
  date         = {1987-06-05},
  doi          = {10.1126/science.236.4806.1287},
  pmid         = {17770330},
  issn         = {0036-8075,1095-9203},
  url          = {http://dx.doi.org/10.1126/science.236.4806.1287},
  urldate      = {2025-11-30},
  language     = {en}
}

@SOFTWARE{YamaneUnknown-xs,
  type     = {software},
  title    = {pmoke - Pulsed {MOKE} Measurement {CLI}},
  author   = {Yamane, Soichiro},
  url      = {https://github.com/Kerr-group/pmoke},
  urldate  = {2025-11-30},
  language = {en}
}

@ARTICLE{Ikeda2024-ja,
  title        = {A concise 40 {T} pulse magnet for condensed matter experiments},
  author       = {Ikeda, Akihiko and Noda, Kosuke and Shimbori, Kotomi and Seki,
                  Kenta and Bhoi, Dilip and Ishita, Azumi and Nakamura, Jin and
                  Matsubayashi, Kazuyuki and Akiba, Kazuto},
  journaltitle = {J. Appl. Phys.},
  publisher    = {AIP Publishing},
  volume       = {136},
  issue        = {17},
  pages        = {175902},
  date         = {2024-11-07},
  doi          = {10.1063/5.0231640},
  issn         = {0021-8979,1089-7550},
  url          = {http://dx.doi.org/10.1063/5.0231640},
  urldate      = {2025-11-30},
  language     = {en}
}

@ARTICLE{Huang2017-ns,
  title        = {Layer-dependent ferromagnetism in a van der Waals crystal down
                  to the monolayer limit},
  author       = {Huang, Bevin and Clark, Genevieve and Navarro-Moratalla, Efrén
                  and Klein, Dahlia R and Cheng, Ran and Seyler, Kyle L and
                  Zhong, Ding and Schmidgall, Emma and McGuire, Michael A and
                  Cobden, David H and Yao, Wang and Xiao, Di and
                  Jarillo-Herrero, Pablo and Xu, Xiaodong},
  journaltitle = {Nature},
  publisher    = {Springer Science and Business Media LLC},
  volume       = {546},
  issue        = {7657},
  pages        = {270--273},
  date         = {2017-06-07},
  doi          = {10.1038/nature22391},
  pmid         = {28593970},
  issn         = {0028-0836,1476-4687},
  url          = {http://dx.doi.org/10.1038/nature22391},
  urldate      = {2025-11-30},
  language     = {en}
}

@ARTICLE{Shimano2013-lc,
  title        = {Quantum Faraday and Kerr rotations in graphene},
  author       = {Shimano, R and Yumoto, G and Yoo, J Y and Matsunaga, R and
                  Tanabe, S and Hibino, H and Morimoto, T and Aoki, H},
  journaltitle = {Nat. Commun.},
  publisher    = {Springer Science and Business Media LLC},
  volume       = {4},
  issue        = {1},
  pages        = {1841},
  date         = {2013-05-14},
  doi          = {10.1038/ncomms2866},
  pmid         = {23673626},
  issn         = {2041-1723,2041-1723},
  url          = {http://dx.doi.org/10.1038/ncomms2866},
  urldate      = {2025-11-30},
  language     = {en}
}

@ARTICLE{Ikeda2025-en,
  title        = {Magneto-optical Kerr-effect measurements under pulsed magnetic
                  fields over 40 {T} using a compact sample fixture},
  author       = {Ikeda, Atsutoshi and Nakamura, Sota and Yamane, Soichiro and
                  Noda, Kosuke and Ikeda, Akihiko and Yonezawa, Shingo},
  journaltitle = {arXiv},
  date         = {2025-11-26},
  eprint       = {2509.07383},
  eprinttype   = {arXiv},
  eprintclass  = {cond-mat.str-el},
  doi          = {10.48550/arXiv.2509.07383},
  addendum = {to be published in Phys.\ Rev.\ Research (DOI: https://doi.org/10.1103/vy7j-ylb4)},
  url          = {http://dx.doi.org/10.48550/arXiv.2509.07383},
  urldate      = {2025-11-30}
}

@ARTICLE{Simsa1980-us,
  title        = {Polar Kerr effect of magnetite in the visible and near
                  infrared spectral regions},
  author       = {Šimša, Z and Legall, H and Široký, P},
  journaltitle = {Phys. Stat. Solidi B Basic Res.},
  publisher    = {Wiley},
  volume       = {100},
  issue        = {2},
  pages        = {665--670},
  date         = {1980-08-01},
  doi          = {10.1002/pssb.2221000234},
  issn         = {1521-3951,0370-1972},
  url          = {https://onlinelibrary.wiley.com/doi/abs/10.1002/pssb.2221000234},
  urldate      = {2025-11-27},
  language     = {en}
}

@ARTICLE{Fontijn1997-hr,
  title        = {Optical and magneto-optical polar Kerr spectra
                  {ofFe3O4andMg2+}- {orAl3+}-{substitutedFe3O4}},
  author       = {Fontijn, W F J and van der Zaag, P J and Devillers, M A C and
                  Brabers, V A M and Metselaar, R},
  journaltitle = {Phys. Rev. B Condens. Matter},
  publisher    = {American Physical Society (APS)},
  volume       = {56},
  issue        = {9},
  pages        = {5432--5442},
  date         = {1997-09-01},
  doi          = {10.1103/physrevb.56.5432},
  issn         = {0163-1829,1095-3795},
  url          = {http://dx.doi.org/10.1103/PhysRevB.56.5432},
  urldate      = {2025-10-13},
  language     = {en}
}

@ARTICLE{Xia2006-rk,
  title        = {High resolution polar Kerr effect measurements of
                  {Sr}$_{2}${RuO}$_{4}$: evidence for broken time-reversal
                  symmetry in the superconducting state},
  author       = {Xia, Jing and Maeno, Yoshiteru and Beyersdorf, Peter T and
                  Fejer, M M and Kapitulnik, Aharon},
  journaltitle = {Phys. Rev. Lett.},
  volume       = {97},
  issue        = {16},
  pages        = {167002},
  date         = {2006-10-20},
  doi          = {10.1103/PhysRevLett.97.167002},
  pmid         = {17155427},
  issn         = {0031-9007},
  url          = {http://dx.doi.org/10.1103/PhysRevLett.97.167002},
  language     = {en}
}

@ARTICLE{Xia2006-sg,
  title        = {Modified Sagnac interferometer for high-sensitivity
                  magneto-optic measurements at cryogenic temperatures},
  author       = {Xia, Jing and Beyersdorf, Peter T and Fejer, M M and
                  Kapitulnik, Aharon},
  journaltitle = {Appl. Phys. Lett.},
  publisher    = {American Institute of Physics},
  volume       = {89},
  issue        = {6},
  pages        = {062508},
  date         = {2006-08-07},
  doi          = {10.1063/1.2336620},
  issn         = {0003-6951},
  url          = {https://doi.org/10.1063/1.2336620}
}

\end{document}